\documentclass[10pt]{revtex4}
\usepackage{amssymb}
\usepackage{latexsym}
\usepackage{epsfig}
\begin{document}
\title{Thermodynamical description of the ghost dark energy model}
\author{M. Honarvaryan$^{1}$, A. Sheykhi$^{1,2}$\footnote{asheykhi@shirazu.ac.ir}and  H. Moradpour $^{1,2}$\footnote{h.moradpour@riaam.ac.ir}}
\address{$^1$ Physics Department and Biruni Observatory, College of
Sciences, Shiraz University, Shiraz 71454, Iran\\
$^2$ Research Institute for Astronomy and Astrophysics of Maragha
(RIAAM), P.O. Box 55134-441, Maragha, Iran}

\begin{abstract}
In this paper, we point out thermodynamical description of ghost
dark energy and its generalization to the early universe.
Thereinafter, we find expressions for the entropy changes of these
dark energy candidates. In addition, considering thermal
fluctuations, thermodynamics of the dark energy component
interacting with a dark matter sector is addressed. {We will also
find the effects of considering the coincidence problem on the
mutual interaction between the dark sectors, and thus the equation
of state parameter of dark energy.} Finally, we derive a relation
between the mutual interaction of the dark components of the
universe, accelerated with the either ghost dark energy or its
generalization, and the thermodynamic fluctuations.
\end{abstract}

\maketitle

\section{Introduction}
{\ B}ig bang, as the cornerstone of standard cosmology, is the
primary generator of the Universe expansion. Bearing the Einstein
equations in mind, the rate of the universe expansion is
determined by a fluid which has significant contribution in
filling the universe, and is called dominated fluid. Since it
seems that the universe is homogeneous and isotropic in scales
larger than $100$ Mpc, the universe expansion is modeled by the
so-called Friedmann-Robertson-Walker (FRW) metric,
\begin{eqnarray}\label{frw}
ds^{2}=dt^{2}-a^{2}\left( t\right) \left[ \frac{dr^{2}}{1-kr^{2}}%
+r^{2}d\Omega ^{2}\right],
\end{eqnarray}
where $a(t)$ is the scale factor, and $k=-1,0,1$ is the curvature
parameter corresponding to open, flat and closed universes
respectively \cite{roos}. It is also shown that the conformal form
of this metric can be used to explain the inhomogeneity of cosmos
in scales smaller than $100$ Mpc \cite{rma}.

Thermodynamic description of the Einstein equations is pointed out
by Jacobson \cite{Jac}. In order to generalize the Jacobson's
results to the cosmological setup, we need to find a causal
boundary for the FRW spacetime. In fact, it was argued that
apparent horizon can be considered as a causal boundary for
dynamical spacetimes \cite{Bak,Hay2,Hay22}. For this propose, the
apparent horizon is defined as the marginally trapped surface
located at
\begin{eqnarray}\label{ah}
\tilde{r}_A=\frac{1}{\sqrt{H^2+{k}/{a^2}}},
\end{eqnarray}
where $H=\dot{a}/{a}$ is the Hubble parameter
\cite{sheyw1,sheyw2}. It was shown that the apparent horizon can
be considered as a causal boundary for the FRW spacetime
associated with gravitational entropy and surface gravity, and
makes the establishment of the  first law of thermodynamics.
\cite{shey2,Cai2,Cai3,CaiKim,Wang,Wangg,Wanggg,Cai4}.

Recent observations indicate that our Universe is currently
undergoing a phase of accelerated expansion with $\dot{a}(t)\geq0$
and $\ddot{a}(t)\geq0$ \cite{Rie,Rie1,Rie2,Rie3}. It implies
either a non-baryonic dominated fluid named dark energy (DE) and
fills about $70$ percent of the universe \cite{rev1,rev2} or
modifying the Einstein equations \cite{meg}. Although the nature
of the dominated fluid needed to support this phase of expansion
is mysterious \cite{roos}, but this phase of expansion is
compatible with the generalized second law of thermodynamics
\cite{pavon1,pavon2,pavon3,msrw}. In fact, the tendency of the
universe to rise its entropy can be interpret as an origin for the
gravity and thus the Einstein equations \cite{verlinde}. This
hypothesis is generalized to various theories of gravity and
cosmological setups
\cite{Vother,Vother1,Vother2,Vother3,Vother4,Vother5,Vother6,
Vother7,Vother8,Vother9,Vother10,Vother11,Vother12,smr,sm}.
Therefore, the tendency of the cosmos, as a closed system, to
increase its entropy is compatible with the current accelerated
phase of the expansion. Finally, we should note that the
thermodynamic analysis of the Universe helps us to get a more
better understanding from the gravity, the Universe, its origin
and evolution, and thus the nature of the mysterious dominated
fluid supporting the current accelerated expansion.

Considering either a new degree of freedom or a new parameter
leads to explanations of DE in the Einstein relativity framework
\cite{de1,de2,de3,de4,de5,de6}. In another approach, based on the
Quantum Chromodynamics (QCD), new particles known as Veneziano
ghost are guaranteed for the DE source \cite{ven}. These particles
explain the $U(1)$ problem in QCD \cite{ven,ven1,ven2}, and yield
a reasonable answer for the time-varying cosmological constant in
a universe with non-trivial geometry \cite{ven3}. The energy
density of the Veneziano ghost DE (GDE) model is proportional to
$\Lambda_{QCD}^3H$, where $\Lambda_{QCD}$ is the QCD mass scale
and thus $\rho_D=\alpha H$ \cite{ven4}. Although the Veneziano GDE
model suffers the coincidence problem, but this shortcoming can be
eliminated by choosing proper value for $\Lambda_{QCD}$
\cite{ven3,ven4}. From thermodynamics point of view, it was shown
that GDE satisfies the generalized second law of thermodynamics
which is in agreement with the current accelerated phase of the
Universe expansion \cite{msrw}. Instability of GDE against
perturbations is studied in \cite{ebsh}. More studies on the GDE
properties can be found in Refs.~\cite{o1,o2,o3,o4,o5,o6,o7,o8}.
In the standard cosmology, dark matter (DM) fills about $26$
percent of the universe which influences the structure formation
and the galaxies rotating curves \cite{roos}. Since the nature of
DM is unknown and its gravitational effects differ from DE, people
usually study DM and DE separately.

{Recently, observational evidences of interaction between DE and
DM are presented \cite{gdo1,gdo2,ob1,ob2,ob3,ob4,ob5,ob6,ob7}.
Moreover, such interactions between the dark sectors of the cosmos
may lead to solve the coincidence problem
\cite{ob7,co1,co2,co3,co4,co5,co6,pavonz}. In addition, bearing
the thermal fluctuation theory in mind \cite{landau}, one can show
that the fluctuation theory may lead to logarithmic correction to
the entropy of event horizon \cite{das}. Its generalization to the
cosmological setup helps us to explain thermal fluctuation of the
cosmos components as the result of the mutual interaction between
the dark sectors of the cosmos
\cite{wangpavon,shs00,shs0,shs,shs1}. Therefore, it seems that DM
should interact with GDE, as a candidate for DE. Such interactions
between DM and GDE are studied in Refs.~\cite{gph,gdm}. Indeed, in
order to solve the coincidence problem, DE should decay into DM
\cite{wangpavon}. Finally, we should note that one may find a
relation for the mutual interaction between the dark sectors by
using the thermal fluctuations of the universe components.}

In general, the vacuum energy of the Veneziano ghost field in QCD
is of the form $H+O(H^2)$ \cite{GGDE}. This indicates that in the
previous works on the GDE model, only the leading term $H$ has
been considered. Indeed, the density of Veneziano ghost is in the
form $\rho_D=\alpha H+\beta H^2$ \cite{GGDE}. It was first pointed
out by Maggiore et al., \cite{GGDE1} that the second term ($\beta
H^2$) might play the crucial role in the early universe when
$\tilde{r}_A\ll1$. Therefore, one should use the generalized
Veneziano ghost density (GGDE) in order to get a more
comprehensive look-out from the effects of considering the
Veneziano ghost as the source of the DE on the universe expansion
and thus related topics. In fact, it was shown that GGDE leads to
better agreement with observational data against GDE \cite{ggde2}.
The GGDE interaction with DM and its relation with observational
data is studied in Ref.~\cite{esa}.

Here, we are going to study thermodynamics of the Universe filled
with GDE/GGDE and DM and in the absence of interaction between DM
and GDE/GGDE. The associated entropy to GDE/GGDE models is
addressed. Thereinafter, we will study thermodynamics of the GDE
and GGDE models when the interaction with DM is considered. We
derive thermodynamic interpretations for such interactions. Since
the WMAP data indicates a flat universe \cite{roos}, we restrict
ourselves to the $k=0$ case. This option leads to $\Omega_k=0$,
where $\Omega_k={k}/({a^2H^2})$ is the dimensionless energy
density parameter induced by the spacetime curvature \cite{roos}.
For the sake of simplicity, we take $c=1$ throughout this paper.

The paper is organized as follows. In the next section, we
consider the GDE model which does not interact with DM and study
its thermodynamics. Section $\textmd{III}$ includes a
thermodynamical description of the GDE interacting with DM. We
investigate the thermodynamics of the non-interacting GGDE in
section $\textmd{IV}$. In section $\textmd{V}$, we provide a
thermodynamical interpretation for the GGDE interacting with DM.
Last section is devoted to summary and concluding remarks.
%%%%%%%%%%%%%%%%%%%%%%%%%%%%%%%%%%%%%%%%%%%%%%%%%%%%%%%%%%%%%%%%%%%%%%%%%%%%%%%%%%%%%%%%
\section{Thermodynamical description of the non-interacting GDE }\label{NonInt}
Here, we consider the flat FRW Universe filled by the GDE and the
pressureless DM, and study the thermodynamics of the GDE whereas
GDE does not interact with DM. For the flat FRW Universe, the
first Friedmann equation is
\begin{eqnarray}\label{fried1}
H^{2}=\frac{1}{3M_{p}^{2}}\left( \rho _{m}+\rho _{D}\right).
\end{eqnarray}
In this equation, $M_p^2=(8 \pi G)^{-1}$ is the reduced Plank
mass, $\rho_m$ is the energy density of the pressureless DM and
$\rho_D$ is the GDE density
\begin{eqnarray}\label{ded}
\rho_{D}=\alpha H,
\end{eqnarray}
where $\alpha$ is a constant of order $\Lambda_{QCD}^3$. With
$\Lambda_{QCD}\sim 100$ MeV and $H\sim 10^{-33}$ eV,
$\Lambda_{QCD}^3H$ gives the right order of magnitude for the
observed DE density \cite{ven4}. For the dimensionless energy
density parameters, we have
\begin{eqnarray}\label{dedp1}
\Omega_{m}&=&\frac{\rho _{m}}{\rho _{cr}}=\frac{\rho _{m}}{
3m_{p^{{}}}^{2}H^{2}}, \\ \nonumber \Omega_{D}&=&\frac{\rho
_{D}}{\rho _{cr}}=\frac{\alpha }{3m_{p}^{2}H}, \label{dedp2}
\end{eqnarray}
where $\rho_c=3M_p^2 H^2$ is critical energy density. Therefore,
the Friedmann equation (\ref{fried1}) can be written as
\begin{eqnarray}\label{fried2}
\Omega_{m}+\Omega_{D}=1.
\end{eqnarray}
Since GDE does not interact with DM, the energy-momentum
conservation implies
\begin{eqnarray}\label{dmc1}
\dot{\rho}_{m}+3H\rho_{m}=0,
\end{eqnarray}
and
\begin{eqnarray}\label{dec1}
\dot{\rho}_{D}+3H\rho_{D}\left( 1+\omega _{D}^{0}\right) =0.
\end{eqnarray}
Here $p_D=\omega _{D}^{0}\rho_D$ where $\omega _{D}^{0}$ is the
equation of state parameter of the non-interacting GDE, and
superscript ($0$) is used to remember that the GDE does not
interact with DM. Taking the derivative of Eqs.~(\ref{fried1})
and~(\ref{ded}) with respect to time, we arrive at
\begin{eqnarray}\label{friedtime}
2H\dot{H}=\frac{1}{3M_{p}^{2}}\left(
\dot{\rho}_{m}+\dot{\rho}_{D}\right),
\end{eqnarray}
and
\begin{eqnarray}\label{gdetime}
\dot{\rho}_{D}=\alpha \dot{H},
\end{eqnarray}
respectively. Combining Eqs.~(\ref{dmc1}),~(\ref{dec1}) with
Eq.~(\ref{friedtime}), we get
\begin{eqnarray}\label{hdot1}
\dot{H}=-\frac{\rho _{D}}{2M_{p}^{2}}\left( 1+u+\omega
_{D}^{0}\right),
\end{eqnarray}
where
\begin{eqnarray}\label{u}
u=\frac{\rho _{m}}{\rho _{D}}=\frac{\Omega _{m}}{\Omega
_{D}}=\frac{1-\Omega _{D}}{\Omega _{D}}.
\end{eqnarray}
Using Eq.~(\ref{hdot1}), one can rewrite Eq.~(\ref{gdetime}) as
\begin{eqnarray}\label{gdetime1}
\dot{\rho}_{D}=\alpha \dot{H}=-\frac{\alpha }{2M_{p}^{2}}\rho
_{D}\left( 1+u+\omega _{D}^{0}\right).
\end{eqnarray}
Inserting this equation into Eq.~(\ref{dec1}), we arrive at
\begin{eqnarray}
-\frac{\alpha }{2M_{p}^{2}}\rho _{D}\left( 1+u+\omega
_{D}^{0}\right) +3H\rho _{D}\left( 1+\omega _{D}^{0}\right)=0,
\end{eqnarray}
which yields
\begin{eqnarray}\label{dec2}
\left( 1+\omega _{D}^{0}\right) \left( 6M_{p}^{2}H-\alpha \right)
=\alpha u.
\end{eqnarray}
Inserting Eq.~(\ref{ded}) into the Eq.~(\ref{fried1}) leads to
\begin{eqnarray}\label{H}
3M_{p}^{2}H=\alpha \left(1+u\right).
\end{eqnarray}
Combining this equation with Eq.~(\ref{dec2}), we get
\begin{eqnarray}
\omega_{D}^{0}=\frac{u}{2u+1}-1.
\end{eqnarray}
Finally, we use Eq.~(\ref{u}) to obtain
\begin{eqnarray}\label{state1}
\omega_{D}^{0}=-\frac{1}{2-\Omega _{D}^{0}},
\end{eqnarray}
Since the apparent horizon is the causal boundary of the FRW
spacetime, we write the first law of thermodynamics on the
apparent horizon in order to find an expression for the entropy of
the GDE as
\begin{eqnarray}\label{flt}
TdS_{D}=dE_{D}+p_{D}dV.
\end{eqnarray}
In this equation, $S_D$ is the entropy associated to the GDE while
$V=\frac{4\pi }{3}\tilde{r}_{A}^{3}$ and $E_D=\rho _{D}V$ are,
respectively, the volume of the flat FRW Universe and the total
energy of the DGE. Note that in flat FRW Universe the apparent
horizon radius is indeed the Hubble radius,
$\tilde{r}_{A}={1}/{H}$. The temperature $T$ of the GDE which will
be equal to the temperature of the apparent horizon when
thermodynamic equilibrium is supposed, and therefore we have
\begin{eqnarray}\label{temp}
 T=\frac{1}{2\pi \tilde{r}_{A}^0}=\frac{H_0}{2\pi }.
\end{eqnarray}
Therefore, for the volume and the total energy we have
\begin{eqnarray}\label{vo}
V=\frac{4\pi }{3}(\tilde{r}_{A}^0)^{3}=\frac{4\pi }{3}H_0^{-3},
\end{eqnarray}
\begin{eqnarray}\label{ener}
E_{D}=\alpha H_0\times \frac{4\pi }{3}H_0^{-3}=\frac{4\pi
}{3}\alpha H_0^{-2}=\frac{4\pi }{3}\alpha (\tilde{r}_{A}^0)^{2}.
\end{eqnarray}
Differentiating relations~(\ref{vo}) and~(\ref{ener}) and
substituting the results into Eq.~(\ref{flt}), we get
\begin{eqnarray}\label{dS0}
dS_{D}^{(0)}=2\pi\tilde{r}_{A}^{0}\left[ \frac{8\pi }{3}\alpha
\tilde{r}_{A}^{0}d\tilde{r}_{A}^{0}+{4\pi\alpha \omega _{D}^{0}}
\tilde{r}_{A}^{0}d\tilde{r}_{A}^{0}\right],
\end{eqnarray}
where we have used $p_{D}=\rho _{D}\omega _{D}^{0}=\alpha
H_{0}\omega _{D}^{0}=\frac{\alpha \omega
_{D}^{0}}{\tilde{r}_{A}^{0}}$. Combining (\ref{dS0})
with~(\ref{state1}), we obtain
\begin{eqnarray}\label{entropygde}
dS_{D}^{\left( 0\right) }=8\pi ^{2}\alpha \left(
\tilde{r}_{A}^{0}\right) ^{2}d\tilde{r}_{A}^{0}\left[
\frac{2}{3}-\frac{1}{2-\Omega _{D}^{0}}\right].
\end{eqnarray}
We should note again that the subscript/superscript ($0$)
indicates that the GDE does not interact with the DM in our model.
Therefore, we find an expression for the entropy changes of the
non-interacting GDE confined by the apparent horizon when the DM
sector does not interact with the GDE sector.
\section{Thermodynamical description of the interacting GDE }\label{Int}
In this section we study the case where the GDE and the
pressureless DM interact with each other. We look at this
interaction as a generator for thermodynamic fluctuations around
equilibrium state. Therefore, bearing the thermodynamic
fluctuations in mind, we try to find an expression for the
interaction. The energy-momentum conservation leads to
\begin{eqnarray}\label{ee}
\dot{\rho}_T+3H(\rho_T+p_T)=0,
\end{eqnarray}
where $\rho_T=\dot{\rho}_m+\dot{\rho}_D$ and $p_T=p_m+p_D=p_D$. In
addition, $H$ is the Hubble parameter of the universe filled with
the interacting GDE, and differs from the $H_0$ introduced in the
previous section. Also, Eq.~(\ref{ah}) is valid for the both of
$H$ and $H_0$ with radii $\tilde{r}_A$ and $\tilde{r}_A^0$,
respectively. Since the dark components interact with each other,
one can decompose Eq.~(\ref{ee}) into the
\begin{eqnarray}\label{emcigde}
\dot{\rho}_{m}+3H\rho _{m}&=&Q, \\
\dot{\rho}_{D}+3H\rho _{D}\left( 1+\omega _{D}^{{}}\right)
&=&-Q,\label{emcigde2}
\end{eqnarray}
Here, $Q=3b^{2}H\left( \rho _{m}+\rho _{D}\right)$ denotes the
interaction term and $b^2$ is a coupling constant \cite{pavonz}.
In addition, $\omega_D=\frac{p_D}{\rho_D}$ is the equation of the
state parameter of the interacting GDE. Using Eq.~(\ref{u}), we
get
\begin{eqnarray}\label{q1}
Q=3b^{2}H\rho_{D}\left(1+u\right).
\end{eqnarray}
Inserting Eqs. (\ref{gdetime1}) and (\ref{q1}) into
Eq.~(\ref{emcigde2}), and doing simple calculations yield
\begin{eqnarray}\label{al}
\alpha \left[ \left( 1+\omega _{D}^{{}}\right) +u\right]
=6M_{p}^{2}H\left[ b^{2}\left( 1+u\right) +\left( 1+\omega
_{D}^{{}}\right) \right].
\end{eqnarray}
Since Eq.~(\ref{H}) is independent of the interaction, we use it
and reach at
\begin{eqnarray}\label{omega}
\omega_{D}=-\frac{2b^{2}\left( 1+u\right) ^{2}+(u+1)}{\left(
2u+1\right)}.
\end{eqnarray}
Using relation (\ref{u}), we get
\begin{eqnarray}\label{omega1}
\omega _{D}=-\frac{1}{2-\Omega _{D}}\left[ \frac{2b^{2}}{\Omega
_{D}}+1 \right],
\end{eqnarray}
as the equation of the state parameter of the interacting GDE.
Inserting Eqs.~(\ref{ded}) and~(\ref{u}) into the Eq.~(\ref{q1}),
we find
\begin{eqnarray}\label{b}
b^{2}=\frac{Q\Omega _{D}}{3\alpha H^{2}},
\end{eqnarray}
which yields
\begin{eqnarray}\label{omega2}
\omega_{D}=-\frac{1}{2-\Omega_{D}}\left[
\frac{2b^{2}}{\Omega_{D}}+1 \right] =-\frac{1}{2-\Omega_{D}}\left[
\frac{2Q}{3\alpha H^{2}}+1\right].
\end{eqnarray}
Bearing the flat FRW spacetime and Eq.~(\ref{ah}) in mind, one can
rewrite this equation as
\begin{eqnarray}\label{omega3}
\omega_{D}=-\frac{1}{2-\Omega_{D}}\left[
\frac{2Q\tilde{r}_{A}^{2}}{3\alpha }+1\right].
\end{eqnarray}
In order to study thermodynamics of interaction term, $Q$, we
consider the first law of thermodynamics
\begin{eqnarray}\label{flt1}
TdS_{D}=dE_{D}+p_{D}dV,
\end{eqnarray}
where $E_D$ and $V_D$ are,  respectively, the total energy and the
volume of the interacting GDE. Here $T$ is the temperature of the
horizon (\ref{temp}) and $p_D$ is the pressure of the interacting
GDE. In addition, $S_{D}$ is the entropy of the interacting GDE
which consists two parts,
\begin{eqnarray}\label{entropygdeinte}
S_{D}=S_{D}^{\left( 0\right) }+S_{D}^{\left( 1\right)},
\end{eqnarray}
where $S_{D}^{\left( 1\right)}=-\frac{1}{2}\ln \left(
CT^{2}\right)$ is a logarithmic correction to the thermodynamic
entropy which is due to the fluctuations around equilibrium, and
is valid in all thermodynamic systems \cite{das}. $C$ is the heat
capacity defined by
\begin{eqnarray}\label{heat}
C=T\frac{\partial S_{D}^{\left( 0\right)}}{\partial
T}=T\frac{\partial S_{D}^{\left( 0\right)}}{\partial
\tilde{r}_{A}^{0}}\frac{\partial \tilde{r}_{A}^{0}}{
\partial T}=T\frac{\partial S_{D}^{\left( 0\right) }}{\partial \tilde{r}_{A}^{0}}
\left(\frac{-1}{2\pi T^{2}}\right).
\end{eqnarray}
It is a matter of calculation to show that
\begin{eqnarray}\label{heat1}
C=-8\pi ^{2}\alpha \left(\tilde{r}_{A}^{0}\right) ^{3}\left[
\frac{2}{3}-\frac{1}{2-\Omega_{D}^{0}}\right]
\end{eqnarray}
and therefore,
\begin{eqnarray}\label{entropy1}
S_{D}^{\left( 1\right) }=-\frac{1}{2}\ln \left[ -2\alpha
\tilde{r}_{A}^{0}\left(
\frac{2}{3}-\frac{1}{2-\Omega_{D}^{0}}\right) \right].
\end{eqnarray}
Using Eq.~(\ref{flt1}) and following the recipe of previous
section, we get
\begin{eqnarray}\label{totalentropy1}
dS_{D}=8\pi ^{2}\alpha \left(\tilde{r}_{A}\right) ^{2}\left(
\frac{2}{3}+\omega_{D}\right)d\tilde{r}_{A},
\end{eqnarray}
as the changes of the total entropy. Combining this equation with
(\ref{omega3}) and using Eq.~(\ref{entropygdeinte}), we obtain
\begin{eqnarray}
Q=\frac{\alpha (2-\Omega
_{D})}{\left(\tilde{r}_{A}\right)^{2}}-\frac{3\alpha
}{2\left(\tilde{r}_{A}\right) ^{2}}-\frac{3\left(
2-\Omega_{D}\right) }{16\pi
^{2}\left(\tilde{r}_{A}\right)^{4}}\left[ \frac{dS_{D}^{\left(
0\right) }}{ d\tilde{r}_{A}}+\frac{dS_{D}^{\left( 1\right)
}}{d\tilde{r}_{A}}\right].
\end{eqnarray}
Since
\begin{eqnarray}
\frac{dS_{D}^{\left( 0\right) }}{d\tilde{r}_{A}}=\frac{\partial
S_{D}^{\left( 0\right) }}{\partial \tilde{r}_{A}^{\left( 0\right)
}}\frac{d\tilde{r}_{A}^{\left(
0\right)}}{d\tilde{r}_{A}}=8\pi^{2}\alpha
\left(\tilde{r}_{A}^{0}\right)^{2}\left( \frac{2}{3}
-\frac{1}{2-\Omega_{D}^{0}}\right) \frac{d\tilde{r}_{A}^{\left(
0\right) }}{d\tilde{r}_{A}},
\end{eqnarray}
and
\begin{eqnarray}
\frac{dS_{D}^{\left( 1\right) }}{d\tilde{r}_{A}}=\frac{\partial
S_{D}^{\left( 1\right) }}{\partial \tilde{r}_{A}^{\left( 0\right)
}}\frac{d\tilde{r}_{A}^{\left(0\right)}}{d\tilde{r}_{A}}=-\frac{1}{2\tilde{r}_{A}^{\left(
0\right)}}\frac{d\tilde{r}_{A}^{\left( 0\right)}}{d\tilde{r}_{A}},
\end{eqnarray}
we reach at
\begin{eqnarray}
Q=\frac{\alpha (2-\Omega_{D})}{\left(\tilde{r}_{A}\right)
^{2}}-\frac{3\alpha }{2\left(\tilde{r}_{A}\right)
^{2}}+\frac{d\tilde{r}_{A}^{\left( 0\right)}}{ d\tilde{r}_{A}}
\left[\frac{3\alpha \left(\tilde{r}_{A}^{0}\right)
^{2}\left(2-\Omega_{D}\right)}{2\left(\tilde{r}_{A}\right)
^{4}(2-\Omega_{D}^{0})}-\frac{\alpha
\left(\tilde{r}_{A}^{0}\right)^{2}\left( 2-\Omega_{D}\right)
}{\left(\tilde{r}_{A}\right) ^{4}}+\frac{3\left( 2-\Omega
_{D}\right) }{32\pi ^{2}\left(\tilde{r}_{A}\right)
^{4}\tilde{r}_{A}^{0}}\right].
\end{eqnarray}
Finally, inserting $\tilde{r}_{A}={1}/{H}$ and
$\tilde{r}_{A}^0={1}/{H_0}$ into this equation, we get
\begin{eqnarray}\label{ff}
Q=\alpha H^{2}(2-\Omega _{D})-\frac{3\alpha
H^2}{2}+\frac{d\tilde{r}_{A}^{\left( 0\right)
}}{d\tilde{r}_{A}}\left[ \frac{3\alpha H^{4}\left( 2-\Omega
_{D}\right) }{2H_{0}^{2}(2-\Omega _{D}^{0})}-\frac{\alpha
H^{4}\left( 2-\Omega _{D}\right)
}{H_{0}^{2}}+\frac{3H^{4}H_{0}\left( 2-\Omega_{D}\right) }{32\pi
^{2}}\right],
\end{eqnarray}
which is an expression for the interaction between the GDE and DM
components of the Universe. {In order to solve the coincidence
problem, DE should decay into the DM sector meaning that $Q>0$
\cite{wangpavon}. Therefore, the permissible thermal fluctuations
are those leading to $Q>0$ and vice versa. The latter means that,
when the mutual interaction between the dark sectors meets the
$Q>0$ condition, then it leaves a physically acceptable
fluctuation into the system. Since we have considered interactions
in the $Q=3b^2H(\rho_D+\rho_m)$ form, the $Q>0$ condition leads to
$b^2>0$. In order to investigate the validity of this criterion,
we use Eq.~(\ref{omega2})
\begin{eqnarray}
\frac{2b^2}{\Omega_D}+1=(\Omega_D-2)\omega_D.
\end{eqnarray}
Because the LHS of this equation is positive and $0<\Omega_D<1$, $\omega_D$ should meet the $\omega_D<0$ condition. Loosely speaking, when the state parameter of DE meets the $\omega_D<0$ condition, its mutual interaction with DM may lead to solve the coincidence problem and induces thermal fluctuations into the system in accordance with Eq.~(\ref{ff}).}
In this way we provide a relation
between the dark components interaction and the thermal
fluctuations around the equilibrium state.
\section{Thermodynamical description of the GGDE } \label{NonIntG}
Here, we consider the flat FRW universe filled by the GGDE and the
DM whereas the dark sectors do not interact with each other.
Therefore, for the energy density of GGDE, we have
\begin{eqnarray}\label{densityGGDE}
\rho _{D}=\alpha H+\beta H^{2},
\end{eqnarray}
where $\beta$ is a free parameter and can be adjusted for better
agreement with observations \cite{ggde2}. In addition, for the
dimensionless density parameter of the GGDE we obtain
\begin{eqnarray}\label{ggdeden}
\Omega_{D}=\frac{\rho_{D}}{\rho_{cr}}=\frac{\left( \alpha +\beta
H\right)}{3m^2_pH},
\end{eqnarray}
where the density parameter of the DM obeys Eq.~(\ref{dedp1}), and
therefore $\Omega_{m}+\Omega_{D}=1$. Since the dark components do
not interact with each other, the energy-momentum conservation
implies
\begin{eqnarray}
&&\dot{\rho}_{m}+3H\rho _{m}=0, \\
&&\dot{\rho}_{D}+3H\rho_{D}\left( 1+\omega _{D}^{0}\right) =0.
\end{eqnarray}
Combining these equations with the time derivative of
Eq.(\ref{fried1}), one obtains
\begin{eqnarray}\label{hdotggde}
\dot{H}=-\frac{\rho_{D}\left( 1+u+\omega_{D}^{0}\right)}{2m_p^2}.
\end{eqnarray}
Now, using the time derivative of Eq.~(\ref{densityGGDE}), we
arrive at
\begin{eqnarray}
\dot{\rho}_{D}=-\frac{\rho_{D}}{2m_p^2}\left( 1+u+\omega
_{D}^{0}\right) \left( \alpha +2\beta H\right),
\end{eqnarray}
which yields
\begin{eqnarray}\label{omega4}
1+\omega_{D}^{0}=\frac{\frac{1}{6m_p^2H}\alpha u+\xi
u}{1-\frac{1}{6m_p^2H}\alpha-\xi},
\end{eqnarray}
and we have defined $\xi={\beta}/{(3m_p^2)}$. If one inserts
Eq.~(\ref{densityGGDE}) into the Eq.~(\ref{fried1}), after using
Eq.~(\ref{u}), then one reaches
\begin{eqnarray}
\frac{\alpha}{6m_p^2H} =\frac{1-\xi \left( u+1\right)}{2\left(
u+1\right)}.
\end{eqnarray}
Using this equation, one can write Eq.~(\ref{omega4}) as
\begin{eqnarray}
1+\omega_{D}^{0}=\frac{\Omega_{D}-\Omega_{D}^{2}+\xi -\xi
\Omega_{D}}{\Omega_{D}(2-\Omega_{D}-\xi )},
\end{eqnarray}
leading to
\begin{eqnarray}\label{omega5}
\omega_{D}^{0}=\frac{\xi
-\Omega_{D}^{0}}{\Omega_{D}^{0}(2-\Omega_{D}^{0}-\xi )},
\end{eqnarray}
which is an expression for the state parameter of the
non-interacting GGDE. The result of the GDE is available in the
$\xi \rightarrow 0$ limit. In order to find the entropy changes of
the GGDE, we assume that the first law of thermodynamics is
available on the apparent horizon, where the pressure of the GGDE
is
\begin{eqnarray}
p_{D}=\rho _{D}\omega _{D^{{}}}^{0}=\left( \alpha H_{0}+\beta
H_{0}^{2}\right) \omega _{D}^{0}=\left( \frac{\alpha
}{\tilde{r}_{A}^{0}}+\frac{ \beta
}{(\tilde{r}_{A}^{0})^{2}}\right) \omega_{D}^{0}.
\end{eqnarray}
and following the recipe of section $\textmd{II}$, we reach
\begin{eqnarray}\label{entropychangesggde}
dS_{D}^{(0)}=8\pi
^{2}(\tilde{r}_{A}^{0})^{2}d\tilde{r}_{A}^{0}\left[
\frac{2}{3}+\frac{\xi-\Omega_{D}^{0}}{\Omega_{D}^{0}(2-\Omega_{D}^{0}-\xi
)}\right] \alpha +8\pi
^{2}\tilde{r}_{A}^{0}d\tilde{r}_{A}^{0}\left[
\frac{1}{3}+\frac{\xi -\Omega_{D}^{0}}{
\Omega_{D}^{0}(2-\Omega_{D}^{0}-\xi )}\right] \beta.
\end{eqnarray}
Again, the subscript/superscript $(0)$ indicates that the GGDE and
the DM do not interact with each other. As a desired result, one
can reach to Eq.~(\ref{entropygde}) by taking appropriate limit
,$\beta \rightarrow 0$, from Eq.~(\ref{entropychangesggde}).
\section{Thermodynamical description of the interacting GGDE }\label{IntG}
Here, based on the thermodynamic fluctuations, we give an
expression for the entropy changes of the interacting GGDE. In
addition, we find a relation between the dark components
interaction and the thermodynamic fluctuations. Since the GGDE and
the DM interact with each other, the energy-momentum conservation
implies
\begin{eqnarray}\label{enerm}
&&\dot{\rho}_{m}+3H\rho _{m}=Q,
\\
&&\dot{\rho}_{D}+3H\rho _{D}\left( 1+\omega _{D}^{{}}\right) =-Q.
\label{enerde}
\end{eqnarray}
Again, $Q=3b^{2}H\left( \rho _{m}+\rho _{D}\right) =3b^{2}H\rho
_{D}\left( 1+u\right)$ is the interaction term, where $b^2$ is a
coupling constant \cite{pavonz}. Combining Eqs.~(\ref{enerm})
and~(\ref{enerde}) with the time derivative of Friedmann
equation~(\ref{friedtime}), and using the time derivative of
Eq.~(\ref{densityGGDE}) leads to
\begin{eqnarray}\label{omega6}
\omega_{D}=-\frac{1}{2-\Omega_{D}-\xi }\left(
1+\frac{2b^{2}}{\Omega_{D}}-\frac{\xi }{\Omega_{D}}\right).
\end{eqnarray}
Now, since $Q=3b^{2}H\rho _{D}\left( 1+u\right)$, we get
\begin{eqnarray}\label{b2}
b^{2}=\frac{Q\Omega _{D}}{3\alpha H^{2}+3\beta H^{3}},
\end{eqnarray}
where we have used Eq.~(\ref{densityGGDE}). Combining this
equation with Eq.~(\ref{omega6}), one obtains
\begin{eqnarray}\label{omegafin}
\omega_{D}=-\frac{1}{2-\Omega_{D}-\xi }\left( 1+\frac{2Q}{(\frac{
3\alpha }{\tilde{r}_{A}^{2}}+\frac{3\beta
}{\tilde{r}_{A}^{3}})}-\frac{\xi }{\Omega_{D}} \right).
\end{eqnarray}
as the relation for the state parameter of the interacting GGDE.
It is obvious that, the results of sections $\textmd{IV}$ and
$\textmd{III}$ are available in the $Q\rightarrow0$ and
$\beta\rightarrow 0$ limits respectively. In addition, the result
of section $\textmd{II}$ is obtainable if one takes the
appropriate limit ($Q\rightarrow0$ and $\beta\rightarrow 0$) from
this equation. Since the mutual interaction between the GGDE and
the DM induces the thermodynamic fluctuations around the
equilibrium, the first law of thermodynamics on the apparent
horizon is available~(\ref{flt}). For the heat capacity we have
\begin{eqnarray}
C=T\frac{\partial S_{D}^{\left( 0\right) }}{\partial T}=T\frac{\partial
S_{D}^{\left( 0\right) }}{\partial \tilde{r}_{A}^{0}}\frac{\partial \tilde{r}_{A}^{0}}{%
\partial T}=T\frac{\partial S_{D}^{\left( 0\right) }}{\partial \tilde{r}_{A}^{0}}%
\left( \frac{-1}{2\pi T^{2}}\right)
=-\tilde{r}_{A}^{0}\frac{\partial S_{D}^{\left( 0\right)
}}{\partial \tilde{r}_{A}^{0}},
\end{eqnarray}
and since
\begin{eqnarray}
\frac{\partial S_{D}^{\left( 0\right) }}{\partial
\tilde{r}_{A}^{0}}=8\pi ^{2}(\tilde{r}_{A}^{0})^{2}\left[
\frac{2}{3}+\frac{\xi -\Omega _{D}^{0}}{\Omega _{D}^{0}(2-\Omega
_{D}^{0}-\xi )}\right] \alpha +8\pi ^{2}\tilde{r}_{A}^{0}\left[
\frac{1}{3}+\frac{\xi -\Omega _{D}^{0}}{\Omega _{D}^{0}(2-\Omega
_{D}^{0}-\xi )}\right] \beta,
\end{eqnarray}
we get
\begin{eqnarray}\label{heat2}
C=-8\pi ^{2}(\tilde{r}_{A}^{0})^{3}\left[ \frac{2}{3}+\frac{\xi
-\Omega _{D}^{0}}{ \Omega _{D}^{0}(2-\Omega _{D}^{0}-\xi )}\right]
\alpha -8\pi ^{2}(\tilde{r}_{A}^{0})^{2}\left[
\frac{1}{3}+\frac{\xi -\Omega _{D}^{0}}{\Omega _{D}^{0}(2-\Omega
_{D}^{0}-\xi )}\right] \beta.
\end{eqnarray}
Therefore, for the logarithmic correction of entropy due to the
thermal fluctuation we reach
\begin{eqnarray}\label{corctentropy}
S_{D}^{\left( 1\right) }=-\frac{1}{2}\ln \left( CT^{2}\right)
=-\frac{1}{2}\ln \left[ -2\tilde{r}_{A}^{0}\left[ \frac{2}{3}
+\frac{\xi -\Omega _{D}^{0}}{\Omega _{D}^{0}(2-\Omega _{D}^{0}-\xi
)}\right] \alpha -2\left[ \frac{1}{3}+\frac{\xi -\Omega
_{D}^{0}}{\Omega _{D}^{0}(2-\Omega _{D}^{0}-\xi )}\right] \beta
\right].
\end{eqnarray}
Considering the first law of thermodynamics and following the
recipe of section $\textmd{IV}$, we obtain
\begin{eqnarray}
dS_{D}=8\pi ^{2}\tilde{r}_{A}^{2}d\tilde{r}_{A}\left[
\frac{2}{3}+\omega _{D} \right] \alpha +8\pi
^{2}\tilde{r}_{A}d\tilde{r}_{A}\left[ \frac{1}{3}+\omega_{D}
\right] \beta,
\end{eqnarray}
where $\tilde{r}_A$ is the apparent horizon radii~(\ref{ah}) of
the universe filled with interacting GGDE and differs from those
of the universe filled by the non-interacting GGDE unless we have
$H=H_0$ which is the non-interaction limit. Now, using
Eq.~(\ref{omegafin}), we get
\begin{eqnarray}\label{entfin}
\frac{dS_{D}}{d\tilde{r}_{A}}=8\pi^{2}\tilde{r}_{A}^{2}\left[
\frac{2}{3}-\frac{1}{2-\Omega _{D}-\xi }\left(
1+\frac{2Q}{\frac{3\alpha }{\tilde{r}_{A}^{2}}+\frac{3\beta }{
\tilde{r}_{A}^{3}}}-\frac{\xi }{\Omega_{D}}\right)
\right]\alpha+8\pi^{2}\tilde{r}_{A}\left[ \frac{1}{3}-\frac{\left(
1+ \frac{2Q}{\frac{3\alpha }{\tilde{r}_{A}^{2}}+\frac{3\beta
}{\tilde{r}_{A}^{3}}}-\frac{\xi }{ \Omega_{D}}\right)}{2-\Omega
_{D}-\xi }\right] \beta.
\end{eqnarray}
Since $\frac{dS_{D}}{d\tilde{r}_{A}}=\frac{dS_{D}^{\left( 0\right)
}}{d\tilde{r}_{A}} +\frac{dS_{D}^{\left(
1\right)}}{d\tilde{r}_{A}}$, we need to evaluate
$\frac{dS_{D}^{\left( 0\right)}}{d\tilde{r}_{A}}$ and
$\frac{dS_{D}^{\left( 1\right)}}{d\tilde{r}_{A}}$. Calculations
lead to
\begin{eqnarray}\label{s0}
\frac{dS_{D}^{\left( 0\right)}}{d\tilde{r}_{A}}=\frac{\partial
S_{D}^{\left( 0\right) }}{\partial \tilde{r}_{A}^{\left( 0\right)
}}\frac{d\tilde{r}_{A}^{\left( 0\right)}}{d\tilde{r}_{A}}=\left[
8\pi ^{2}(\tilde{r}_{A}^{0})^{2}\left[\frac{2}{3}+\frac{\xi
-\Omega _{D}^{0}}{\Omega _{D}^{0}(2-\Omega_{D}^{0}-\xi )}\right]
\alpha +8\pi ^{2}\tilde{r}_{A}^{0}\left[ \frac{1}{3}+\frac{\xi
-\Omega _{D}^{0}}{\Omega_{D}^{0}(2-\Omega_{D}^{0}-\xi )}\right]
\beta \right] \frac{d\tilde{r}_{A}^{\left(0\right)
}}{d\tilde{r}_{A}}
\end{eqnarray}
and
\begin{eqnarray}\label{s1}
\frac{dS_{D}^{\left( 1\right)}}{d\tilde{r}_{A}}=\frac{\partial
S_{D}^{\left(1\right)}}{\partial \tilde{r}_{A}^{\left( 0\right)
}}\frac{d\tilde{r}_{A}^{\left( 0\right)}
}{d\tilde{r}_{A}}=\frac{-1}{2\tilde{r}_{A}^{\left( 0\right)
}}\frac{d\tilde{r}_{A}^{\left( 0\right) }}{d\tilde{r}_{A}}.
\end{eqnarray}
Finally, we get
\begin{eqnarray}
\frac{dS_{D}^{\left( 0\right)
}}{d\tilde{r}_{A}}+\frac{dS_{D}^{\left(
1\right)}}{d\tilde{r}_{A}}=\left[ 8\pi
^{2}(\tilde{r}_{A}^{0})^{2}\left[ \frac{2}{3}+\frac{\xi -\Omega
_{D}^{0}}{\Omega_{D}^{0}(2-\Omega _{D}^{0}-\xi )}\right] \alpha
+8\pi^{2}\tilde{r}_{A}^{0}\left[\frac{1}{3}+\frac{\xi -\Omega
_{D}^{0}}{\Omega_{D}^{0}(2-\Omega _{D}^{0}-\xi )}\right] \beta
-\frac{1}{2\tilde{r}_{A}^{\left( 0\right) }}\right]
\frac{d\tilde{r}_{A}^{\left( 0\right) }}{d\tilde{r}_{A}}.
\end{eqnarray}
Combining this equation with Eq.~(\ref{entfin}), we reach
\begin{eqnarray}\label{final}
&Q&=\frac{3}{2\tilde{r}_{A}^4}[\alpha([\frac{2}{3}+\frac{\xi-\Omega}{\Omega(2-\Omega-\xi)}]\tilde{r}_{A}^2-
(\tilde{r}_{A}^{\left(0\right)})^2[\frac{2}{3}+\frac{\xi-\Omega^0}{\Omega^0(2-\Omega^0-\xi)}]
\frac{d\tilde{r}_{A}^{\left(0\right)}}{d\tilde{r}_{A}})\nonumber \\
&+&
\beta([\frac{1}{3}+\frac{\xi-\Omega}{\Omega(2-\Omega-\xi)}]\tilde{r}_{A}-
\tilde{r}_{A}^{\left(
0\right)}[\frac{1}{3}+\frac{\xi-\Omega^0}{\Omega^0(2-\Omega^0-\xi)}]\frac{d\tilde{r}_{A}^{\left(
0\right) }}{d\tilde{r}_{A}})]+\frac{3}{4\tilde{r}_{A}^{\left(
0\right) }\tilde{r}_{A}^4}\frac{d\tilde{r}_{A}^{\left( 0\right)
}}{d\tilde{r}_{A}}.
\end{eqnarray}
Again, $\tilde{r}_{A}^{\left( 0\right) }$ is the radii of the
apparent horizon when the GGDE does not interact with the DM and
therefore, there is no thermal fluctuations around the equilibrium
state of the universe. In addition, $\tilde{r}_{A}$ is the radii
of the universe accelerated by an interacting GGDE. {In order to
alleviate the coincidence problem we should have $b^2>0$. From
Eq.~(\ref{omega}) we get
\begin{eqnarray}
-\left(\frac{2b^2}{\Omega_D}+1\right)=\omega_D(2-\Omega_D-\xi)-\frac{\xi}{\Omega_D}.
\end{eqnarray}
Since the LHS of this equation is negative, we should have
$\omega_D(2-\Omega_D-\xi)-\frac{\xi}{\Omega_D}<0$ leading to the
$\omega_D<\frac{1}{\Omega_D(\frac{2}{\xi}-\frac{\Omega_D}{\xi}-1)}$
condition for $\omega_D$. Therefore, when the equation of state
parameter of DE meets the
$\omega_D<\frac{1}{\Omega_D(\frac{2}{\xi}-\frac{\Omega_D}{\xi}-1)}$
condition, DE decays into DM in agreement with the solving of the
coincidence problem leading to leave thermal fluctuations in the
system.} Finally, we should note that the Eq.~(\ref{final}) is
indeed, a relation between the interaction and the thermodynamic
fluctuations.
%%%%%%%%%%%%%%%%%%%%%%%%%%%%%%%%%%%%%%%%%%%%%%%%%%%%%%%%%%%%%%%%%%%%%%%%%%%%%%%%%%%%%
\section{Summary and discussion}\label{CONC}
We have investigated thermodynamics of the GDE enclosed by the
apparent horizon in the flat FRW Universe, and found expressions
for its entropy as well as the equation of state parameter. Since
some observation data points to the mutual interaction between the
dark components of the Universe, we have studied thermodynamics of
the interacting GDE. Indeed, such interactions may lead to
perturbations in the Hubble parameter yielding different radii and
temperature for the apparent horizon. Therefore, it seems that one
can consider such interactions as the cause of the fluctuations
around the equilibrium state which is the state of the Universe
when the dark sectors do not interact with each other. In
addition, we obtained an expression for the equation of state
parameter of the interacting GDE, as  well as an expression for
its entropy changes. Bearing the logarithmic correction to the
entropy in mind, we got a relation between the mutual interaction
of the dark components of the Universe and the thermal
fluctuations around the equilibrium. In order to study a more
realistic model, we pointed to the GGDE model where its energy
density profile consists two parts including the term proportional
to $H$ and subleading term which is proportional to $H^2$. This
subleading term may play the key role in the early Universe
\cite{ggde2}. We investigated thermodynamics of GGDE, and found
expressions for its entropy changes and the equation of state
parameter. In continue, we got an expression for the equation of
state parameter of the interacting GGDE as well as a relation
between the thermal fluctuations and the mutual interaction of the
GGDE and the pressureless DM. {Finally, we showed that the
coincidence problem makes a limitation on the possible mutual
interaction between the dark sectors and thus the state parameter
of the DE candidates. The corresponding limitations on the state
parameter were derived for both of the GDE and GGDE models.}
Briefly, we think our survey shows that the interaction between
the dark sectors of the Universe can be considered as the cause of
the thermal fluctuations and thus, the logarithmic correction to
the entropy of the GDE and GGDE models.
%%%%%%%%%%%%%%%%%%%%%%%%%%%%%%%%%%%%%%%%%%%%%%%%%%%%%%%%%%%%%%%%%%%%%%%%%%%%%
\acknowledgments{A. Sheykhi thanks Shiraz University Research
Council. This work has been supported financially by Research
Institute for Astronomy and Astrophysics of Maragha, Iran.}
%%%%%%%%%%%%%%%%%%%%%%%%%%%%%%%%%%%%%%%%%%%%%%%%%%%%%%%%%%%%%%%%%%

\end{document}